\documentclass[conference]{IEEEtran}
\usepackage{cite}
\usepackage{amsmath,amssymb,amsfonts}
\usepackage{algorithmic}
\usepackage{graphicx}
\usepackage{textcomp}
\usepackage{xcolor}
\usepackage{hyperref}
\usepackage{url}
\usepackage{todonotes}
\usepackage[utf8]{inputenc}
\usepackage[T1]{fontenc}
\usepackage[english]{babel}
\usepackage{subcaption}
\usepackage{changes}
\newcommand\tab[1][1cm]{\hspace*{#1}}
\newcommand\blfootnote[1]{%
  \begingroup
  \renewcommand\thefootnote{}\footnotetext{#1}%
  \endgroup
}

\def\BibTeX{{\rm B\kern-.05em{\sc i\kern-.025em b}\kern-.08em
    T\kern-.1667em\lower.7ex\hbox{E}\kern-.125emX}}
\begin{document}

\title{Methods to Increase the Amount of Data for Speech Recognition for Low Resource Languages\\
}

\author{\IEEEauthorblockN{\textsuperscript{*}Alexan Ayrapetyan}
\IEEEauthorblockA{\textit{NVIDIA}\\
Yerevan, Armenia \\
aayrapetryan@nvidia.com}
\and
\IEEEauthorblockN{\textsuperscript{*}Sofia Kostandian}
\IEEEauthorblockA{\textit{NVIDIA}\\
Yerevan, Armenia\\
skostandian@nvidia.com}
\and
\IEEEauthorblockN{\textsuperscript{*}Ara Yeroyan}
\IEEEauthorblockA{\textit{Plat.ai}\\
Yerevan, Armenia  \\
0009-0009-3605-9702}
\and
\IEEEauthorblockN{\textsuperscript{*}Mher Yerznkanyan}
\IEEEauthorblockA{\textit{Buymie}\\
Yerevan, Armenia \\
myerznkanyan@gmail.com}
\and

\IEEEauthorblockN{Nikolay Karpov}
\IEEEauthorblockA{\textit{NVIDIA}\\
Yerevan, Armenia \\
nkarpov@nvidia.com}
\and
\IEEEauthorblockN{Nune Tadevosyan}
\IEEEauthorblockA{\textit{NVIDIA}\\
Yerevan, Armenia \\
ntadevosyan@nvidia.com}
\and
\IEEEauthorblockN{Vitaly Lavrukhin}
\IEEEauthorblockA{\textit{NVIDIA}\\
Santa Clara, US \\
vlavrukhin@nvidia.com}
\and
\IEEEauthorblockN{Boris Ginsburg}
\IEEEauthorblockA{\textit{NVIDIA}\\
Santa Clara, US \\
}
}
\maketitle

\begin{abstract}
This study explores methods to increase data volume for low-resource languages using techniques such as crowd-sourcing, pseudo-labeling, advanced data preprocessing and various permissive data sources such as audiobooks, Common Voice, YouTube. While these methods are well-explored for high-resource languages, their application for low-resource languages remains underexplored. Using Armenian and Georgian as case studies, we demonstrate how linguistic and resource-specific characteristics influence the success of these methods.

This work provides practical guidance for researchers to choose cost-effective and quality-driven dataset extension strategies for low-resource languages.
The key takeaway from various data extension approaches is that paid crowd-sourcing offers the best balance between cost and quality, outperforming volunteer crowd-sourcing, open-source audiobooks, and unlabeled data usage.

Ablation study shows that models trained on the expanded datasets outperform existing baselines and achieve 5.73\% for Gergian and 9.9\% for Armenian ASR word error rate using a relatively small FastConformer architecture.
We open-sourced both the Armenian\footnote{\url{https://huggingface.co/nvidia/stt_hy_fastconformer_hybrid_large_pc}} and Georgian\footnote{\url{https://huggingface.co/nvidia/stt_ka_fastconformer_hybrid_large_pc}} models to allow further research and practical applications.

\end{abstract}


\section{Introduction}

\blfootnote{* Equal contribution}

Armenian and Georgian, despite their unique linguistic features, are considered low-resource languages due to their limited digital presence and small speaker populations. These factors limit efforts to increase the amount of ASR data in various ways. To address this, it is necessary to identify an optimal strategy to overcome these challenges. During our research, we found that the amount of data available for Georgian, including unvalidated data, is much larger than what is available for Armenian. Our experiments showed that the Georgian data is enough to train a reasonably good ASR model, but the available Armenian data is still too small to produce a model of the same quality.

The objective of this study is to explore methods for expanding data collection using all available sources for the Armenian language, aiming to match the amount of labeled data available for Georgian. This work involves not only creating an open data corpus for training but also evaluating methods for data collection in terms of their impact on the quality of Automatic Speech Recognition (ASR) models. The investigation also includes a comprehensive assessment of the impact of these methods on ASR model performance, offering practical insights into their effectiveness.

The main contributions are the following:
\begin{itemize}
    \item Increased the open Common Voice dataset's size by 10 times (from 5 to 50 hours).
    \item Prepared a new open audio-book ASR dataset.\footnote{https://www.openslr.org/154}
    \item Developed a new open-text corpus with modern Armenian language.\footnote{https://www.openslr.org/153}
    \item Created a new crowd-sourced speech dataset using developed text corpus.
    \item Extracted a new unlabeled audio dataset useful for self-training.
    \item Analyzed the effectiveness of additional data for improving ASR model quality.
    \item Established Armenian and Georgian data processing pipelines for each dataset.
    \item Achieved SOTA ASR performance for low-resource languages using a relatively small FastConformer model and disseminated them under a production-friendly license.\footnote{\href{https://huggingface.co/nvidia}{https://huggingface.co/nvidia}}
\end{itemize}

This work not only contributes for Armenian and Georgian dataset development but also provides valuable experience for scaling any low-resource ASR dataset in other low-resource languages by systematically evaluating diverse data collection strategies and their impact on ASR quality.

\section{Related work}

Large-scale projects have included datasets even for low-resource languages, such as FLEURS (Few-shot Learning Evaluation of Universal Representations of Speech) \cite{Conneau2022FLEURS:} and  Common Voice\footnote{\href{https://commonvoice.mozilla.org/en}https://commonvoice.mozilla.org/en} \cite{Ardila2019CommonVoice}. We have narrow focus just two languages Georgian and Armenian.

Previous study on  minority languages \cite{bartelds2023making} demonstrates notable boost in ASR performance after data augmentation, self-training, and utilization of a pre-existing text-to-speech (TTS) system. Our work differs from it in that we focused on extension of amount of real-life data. 
Bhogale et al. \cite{bhogale2024empowering} have also used, unlabeled data to improve ASR performance in low-resource language. While they mainly rely on pseudo-labeling, we use it as one of several methods for dataset expansion.

Several studies have addressed Georgian ASR, including experimental models available on Hugging Face.
In \cite{laptev2021dynamic}, the authors tackle low-resource ASR challenges by introducing dynamic augmentation and BPE-Dropout strategies, achieving 31.1 WER and 17.2 CER for Georgian. Another study \cite{romanenko2022robust} explores data augmentation, model adaptation, and multi-task learning for low-resource languages, with WER results ranging between 34.03 and 34.6. Additionally, \cite{kardava2016georgian} examines the implementation of Georgian ASR in popular search engines, highlighting challenges and successes in practical applications.
    
    Additionally, there have been various tests of different Hugging Face based open-source Georgian ASR models on Mozilla Common Voice (MCV) data. For instance:
        \begin{enumerate}
            \item Wav2Vec based checkpoints: Results (WER) include 43.860\footnote{\href{https://huggingface.co/m3hrdadfi/wav2vec2-large-xlsr-georgian}{wav2vec2-large-xlsr-georgian}}, 48.340\footnote{\href{https://huggingface.co/Temur/wav2vec2-Georgian-Daytona}{wav2vec2-Georgian-Daytona}}, and 15.270\footnote{\href{https://huggingface.co/arampacha/wav2vec2-xls-r-1b-ka}{wav2vec2-xls-r-1b-ka}}.
            \item Whisper based checkpoints: Results (WER) include 43.173\footnote{\href{https://huggingface.co/GreatSarmad/whisper-small-ka}{whisper-small-ka}}, 31.855\footnote{\href{https://huggingface.co/anuragshas/whisper-large-v2-ka}{whisper-large-v2-ka}}.
        \end{enumerate}
Based on these experiments and results, it is evident that the outcomes are approximately similar across different studies and experiments. Therefore, we decided not to rely on existing models and instead aim to develop new Georgian models, experiments, and results.

Before end2end networks, a comprehensive method for enhancing Armenian speech recognition was developed using the CMUSphinx toolkit \cite{vardanyan2016noise}. This work addressed the acoustic challenges posed by background noise, specifically targeting the Armenian language, a representative of low-resource languages. 
Another approach \cite{baghdasaryan2022armspeech} developed the ArmSpeech corpus. The corpus assembly involved sourcing high-quality audio samples from freely available audiobooks and real-life speech scenarios, resulting in a total of 11.77 hours of data. The corpus is not open-source and remains private.

A recent study \cite{yeroyan2024enabling} demonstrated an effective method for creating ASR datasets from Armenian audiobooks. Their approach utilized a pipeline VAD-ASR-CER (VAC) to align and segment long-duration audio into manageable datasets suitable for ASR training. We used this approach to prepare our new audiobooks' dataset.

\section{Dataset Extension}

\subsection{Common Voice Offline Events}

When we started developing Armenian ASR, the Common Voice corpus had just 5 hours of Armenian data, with only 3 hours validated. To address this, we first increased available texts by adding old open-license books and then organized two data collection events at the two largest universities in Armenia. 
Participants were invited to read aloud and record texts from Common Voice, or to validate the labeled data submitted by other contributors. To encourage participation, we offered certificates and exclusive custom merchandise.

During the two events an additional 48 hours of speech data were collected, and 47 hours were validated. We spent about \$1000 to buy merchandise and about a month of a manager's work to prepare.
These offline events were costly but are likely the most sustainable way to grow the dataset, as they raised awareness of the Common Voice platform for future contributions.

\subsection{Crowd-sourcing}

Instead of relying on outdated sources, we have partnered with Yerevan City Magazine, which agreed that we share their texts as part of the Yerevan Magazine Open Texts collection\footnote{https://www.openslr.org/153/}.

We processed these texts and used the Toloka\footnote{https://toloka.ai/}  platform for crowd-sourced reading. Following a method from earlier research\cite{karpov21_interspeech} and implemented start, validate, and download stages using SDP pipelines\footnote{https://github.com/NVIDIA/NeMo-speech-data-processor/pull/64}.  Automatic validation was performed with our ASR model, trained on MCV and Fleurs datasets. The entire process took two months cost, and around \$150, and one month of engineering work is required one month of engineering work. Ultimately, this cost-effective approach added 69.23 hours of verified speech data to our dataset.

\subsection{Audiobooks}
\label{sec:audiobooks_intro}

The use of audiobooks as a means to extend our dataset is inspired by their success in a major speech recognition corpora such as LibriSpeech, which originates from the Gutenberg Project \cite{stroube2003literary}. However, there are inherent challenges associated with audiobooks as they feature lengthy audio chapters, in contrast to ASR preferred 3 to 15 seconds long audio segments. Inspired by \cite{yeroyan2024enabling}, we used their open-source package vac\_aligner\footnote{VAC's package - https://pypi.org/project/vac-aligner/} to align and segment the long-duration Grqaser's\footnote{A digital library of Armenian audiobooks - https://grqaser.org} audiobooks into manageable segments suitable for ASR training. This method enabled us to create an additional 20-hour corpus\footnote{https://www.openslr.org/154} for the Armenian speech, costing about one month of scientist's work.

\subsection{Pseudo-labeling}
\label{sec:pseudo}
To enhance the diversity and volume of speech data for ASR training, we identified suitable public Armenian audio content on YouTube. We curated high-quality data, focusing on interviews, shows, and podcasts for their clarity, minimal noise distributed under a permissive Creative Commons (CC BY) license. 

NeMo’s speech preprocessing tools structured our workflow. An initial manifest listed audio files, followed by quality checks: language identification to remove non-Armenian content and voice activity detection (VAD) using the \text{vad\_multilingual\_frame\_marblenet} model to isolate speech segments. Standardization ensured consistent filenames and metadata, and segment durations were set to 20 seconds to avoid splitting words.
We then used NeMo’s force aligner and base model for word-level predictions and timestamps, facilitating segmentation into 20-second chunks with intact words. A model removed silent segments, retaining only relevant speech data.

\begin{table}[h!]
\centering
\begin{tabular}{|l|l|l|r|r|}
\hline
\textbf{Corpus}        & \textbf{Executor}  & \textbf{Time Spent} & \textbf{Extra \$}  & \textbf{Output Hours}  \\ \hline
Common Voice & Organizer & 1 month & 1000 &  43.2\\ \hline
Crowd & Engineer & 1 month  & 150 &  69.23  \\ \hline
Audiobooks & Scientist  & 1 month & 0 &  21.96   \\ \hline
YouTube & Engineer     & 1 month & 0 & unlabeled 145  \\ \hline
Total & \multicolumn{2}{c|}{4 man-months} & 1150 & 279.39   \\ \hline
\end{tabular}
\caption{Costs and work results analysis for different corpora.}
\label{costs}
\end{table}

This process yielded a cleaned dataset of approximately 145 hours useful for iterative pseudo-labeling (IPL) \cite{likhomanenko2020slimipl}, costing us about one month of an engineer.

\bigskip

\section{Training Setup}

    Training our models varies based on how we handle punctuation and capitalization:

    \begin{itemize}
        \item Approach 1: Train with punctuation and capitalization (PC). For Georgian, we keep periods, commas, and question marks; for Armenian, we retain all punctuation.\footnote{\url{https://en.wikipedia.org/wiki/Armenian_(Unicode_block)}}

        \item Approach 2: Train without punctuation and capitalization (no PC). All punctuation is removed, and letters are treated uniformly, without distinguishing between uppercase and lowercase.
    \end{itemize}

    It is important to note that Georgian is a unicameral language, meaning it does not have uppercase or lowercase letters. For this reason, the training process retains punctuation, as in the first method while processing letters uniformly, as in the second. To ensure reliable evaluation and avoid inflated performance metrics, we carefully preprocess all datasets by removing any overlap between training and testing splits.

\begin{table*}[t]
    \centering
    \resizebox{1.0\textwidth}{!}{
    \begin{tabular}{|l|c|c|c|c|c|c|}
        \hline  
        \tab[0.5cm] \textbf{Test Subset Metric} & \multicolumn{2}{c|}{\textbf{MCV WER (\%)}}  & \textbf{MCV PER (\%)} & \multicolumn{2}{c|}{\textbf{FLEURS WER (\%)}} & \textbf{FLEURS PER (\%)} \\
        \textbf{Train Subset} & \textbf{ with P } & \textbf{w/o P} &  & \textbf{with P} & \textbf{w/o P} &   \\
        \hline
        MCV FLEURS & 5.73  & 4.35 & 9.87  & 13.44  & 10.85 & 27.98 \\
        \hline \hline
        Whisper Large V3 & 78.16 & 76.82 & 48.92  & 89.21 & 88.31  & 89.21  \\
        \hline 
        Seamless & 11.14 &  9.52 & 9.56 & 13.57 & 9.79 & 22.7   \\
        \hline
    \end{tabular}
    }
    \caption{WER and PER values for the best our Georgian model (RNN-T head) with and without punctuation. }
    \label{tab:results_ge}
\end{table*}



All final models were trained with modified parameters from the standard configuration file\footnote{\href{https://github.com/NVIDIA/NeMo/blob/main/examples/asr/conf/fastconformer/hybrid_transducer_ctc}{hybrid\_transducer\_ctc}}. Settings were adapted for both languages to improve the training process and achieve optimal results.
After training, we average the weights of the five best checkpoints. We measure model quality using Word Error Rate (WER) and Punctuation Error Rate (PER) \cite{meister2023librispeech}.

The initial model, pre-trained on English, is FastConformer Hybrid Large, which combines both RNN-T and CTC decoders, where the RNN-T head typically outperforms CTC as it is more context-aware and better suited for handling real-world speech data. We selected FastConformer Hybrid because the models we compared it to, Whisper and Seamless, are considerably larger. For low-resource languages, using such large models risks overfitting and inefficient resource use during both training and inference. Since our goal is to develop small, fast, and accurate models, FastConformer Hybrid was the most practical choice. Key training parameters are 1024 global batch size, 40 warmup steps (ratio 0.1), an initial learning rate of 0.005-0.006 with a cosine annealing scheduler, and training across 8 GPUs for 150-200 epochs.


\section{Experiment Results}

In our experiments, we achieved a WER of 5.73\% for Georgian, indicating the sufficiency of the available dataset for proper modeling. Therefore, we decided to proceed with the existing data, which included the full MCV v.17 dataset, which included 76.38 hours of validated training, 19.82 hours of validated development, and 63.4 hours of not validated (Other) data. Additionally, we included 3 hours of FLEURS training and 0.8 hours of FLEURS development data. Our results (\ref{tab:results_ge}), including comparisons with Seamless and Whisper, show that our model outperformed both, achieving lower WER and higher accuracy across most test sets. Thus, our approach works well for improving ASR performance for Georgian using a mix of validated and non-validated data. 

For Armenian, Table \ref{costs} shows that we successfully achieved our goal of increasing its labeled dataset to match that of the Georgian. We also conducted an ablation study to evaluate the significance of each component in relation to its associated costs.
For each pair of WER at the Table\ref{tab:results_am} we estimated a confidence interval and counted the probability of improvement (POI) defined by M. Bisani \cite{bisani2004bootstrap}. All POI values were above the 95\% threshold except for pair (\ref{tab:results_am} rows 7 vs 10) where POI is about 0.85, which is still acceptable.

\begin{table*}[t]
    \centering
    \small
    \resizebox{1.0\textwidth}{!}{
    \begin{tabular}{|c|l|c|ll|l|ll|l|}
    \hline
    \textbf{\#} & \tab[1.9cm]  \textbf{Test Subset Metric} &  &  \multicolumn{2}{c|}{\textbf{MCV WER (\%)}} & \textbf{MCV PER (\%)} &  \multicolumn{2}{c|}{\textbf{FLEURS WER (\%)}} & \textbf{FLEURS PER (\%)} \\ 
    & \textbf{Train Labeled Subset  } & \textbf{Unlabeled} & \textbf{with PC} & \textbf{without PC} & \textbf{} & \textbf{with PC} & \textbf{without PC} & \textbf{} \\ \hline
    1 & MCV FLEURS (no PC)  & & -  & 10.93 & - & -  & 10.80 &  - \\ \hline
    2 & MCV FLEURS (no PC) & YouTube & -  & 8.98 (-17\%) & - & -  & 7.41 (-31\%) & - \\ 
    \hline \hline
    3 & MCV FLEURS  & & 13.41  & 10.59   & 15.48   & 19.26   & 13.14  & 15.64 \\ \hline
    4 & MCV FLEURS & YouTube & 12.81 (-4\%) & 9.85 (-7\%) & 15.38 (-1\%) & 16.17 (-16\%) & 9.7 (-26\%) & 17.05 (+16\%) \\     \hline
    5 & MCV FLEURS Crowd & & 10.83 (-19\%)  & 8.04 (-24\%) & 15.78 (+2\%) & 14.56 (-24\%) & 8.80 (-33\%) & 16.01 (+2\%) \\ \hline
    6 & MCV FLEURS Crowd & YouTube & 10.62 (-20\%)  & 9.03 (-15\%) & 15.63 (+1\%) & 13.67 (-29\%) & 8.17 (-38\%)  & 15.9 (+2\%) \\     \hline
    7 & MCV FLEURS Crowd Audiobooks & & 10.8 (-19\%) & 7.39 (-30\%) & \textbf{14.96 (-2\%)} & 13.03 (-32\%) & 7.40 (-44\%) & \textbf{15.54 (-1\%)} \\ \hline
    8 & MCV FLEURS Crowd Audiobooks & YouTube & \textbf{9.90 (-26\%) } & \textbf{7.09 (-33\%)} & 15.68 (+1\%) & \textbf{12.32 (-36\%)} & \textbf{ 6.72 (-49\%)} & 15.95 (+2\%)\\  \hline
    & & Audiobooks  &  & & & & & \\
    9 & MCV FLEURS Crowd & YouTube & 10.49 (-22\%)  & 7.61 (-28\%) & 15.86 (+2\%) & 12.56 (-35\%) & 6.86 (-47\%) & 16.06 (+3\%) \\     \hline
    \hline 
    10 & Whisper Large v3 & & 54.17 & 50.84  & 23.32 & 45.60 & 40.75 & 33.62 \\ \hline
    11 & Seamless & & 14.43 & 6.46 & 16.41 & 12.74 & 7.54 & 26.19 \\ \hline
    \end{tabular}
     }
    \caption{WER and PER values for Armenian models (RNN-T head) with PC and without PC.}
    \label{tab:results_am}
\end{table*}

\subsection{Common Voice}

We started by evaluating two proprietary models, Whisper Large v3 and Seamless without any fine-tuning, to establish baselines, revealing the limitations of current methods for low-resource languages. Then, we developed models using the MCV (43.2 hours) and FLEURS datasets with and without PC, setting benchmarks that highlighted areas needing improvement.
The model trained with PC and evaluated without PC shows slightly better WER (10.59\% vs 10.93\%) than the model trained without PC on a relatively simple MCV test. 
However, a more complicated FLEURS test set demonstrated reverse behavior. 
Anyway, we decided that the model is capable enough to catch information about PC and trained further models on the extended datasets only with PC.

\subsection{Crowd Sourcing}

The Crowd-sourced part is almost 70 hours of data which is the biggest part of our labeled set. Adding it to the train set improved WER for the MCV test with punctuation and capitalization from 13.41 to 10.83, which is a 19\% relative improvement, and the FLEURS test from 19.26 to 14.56 (24\% relatively). These are the best relative improvements in our ablation study. We explain this not only by the significant size of data but also by the good quality of speech achieved by modern text usage from Yerevan City Magazine.

\subsection{Audiobooks}
\label{sec:audiobooks_experiments}

Audiobooks, as described in Section \ref{sec:audiobooks_intro}, tend to have longer, continuous speech with varied prosody, intonation and speaking styles. When combining audiobooks with only MCV dataset, the overall performance of the model degrades (shown in \cite{yeroyan2024enabling}). To validate the "credibility" of source data, we trained two models - one with using the source labels (texts) and the other using the pseudo-labelling \cite{likhomanenko2020slimipl} (omitting the true labels). 
Using a supervised version of Audiobooks (\ref{tab:results_am} row 9) we get superior results on all model-test combinations than with semi-supervised (\ref{tab:results_am} row 10) setup. This might seem to contradict to \cite{yeroyan2024enabling}'s results, which showed worsened WER on MCV test when using mixed  MCV and Audiobooks train data, but there is a big catch. In our experiments we used much bigger training corpus (extra two sources - FLEURS and Crowd), demonstrating that Audiobooks themselves are a valid source; just need an appropriate amount of mixed data to reach better generalization (adding Audiobooks helped to decrease the test WER on MCV data).

For a better illustration, we also conducted experiments with vs without using Audiobooks. 
Looking into Table \ref{tab:results_am} row 9: where train data is a mix of [\textbf{MCV FLEURS Crowd Audiobooks} + \textbf{YouTube} (pseudo-labeled)], where we added Audiobooks on top of the training dataset used in - Table \ref{tab:results_am} row 7: \textbf{MCV FLEURS Crowd} + \textbf{YouTube} - we observe a notable enhancement evidenced by a significant 21\%  decrease from 10.62\% to 9.9\% WER on the MCV and a 17\% decrease (from 13.67\% to 12.32\%  WER) on the FLEURS test sets. Other model / test set combinations of those experiments show similar progress, too. Hence, the Audiobooks helped to improve the results on MCV test (dataset from other domain), yielding the best results among all the experiments, through a better model generalization. Therefore, we helped the authors from \href{www.grqaser.org}{Grqaser} to make the constructed dataset (21.96 hours of audio-text pairs) accessible to the general public via publishing in OpenSLR.

\subsection{Pseudo-labeling}

We applied four experiments to explore how iterative pseudo-labeling algorithm with unlabelled data could address the lack of data. 
The first model trained on \textbf{MCV} and \textbf{FLEURS (no PC)} decreased MCV WER by 17\% and FLEURS by 31\% after adding unlabeled YouTube dataset. 
Next, three experiments we perform with PC utilizing the same amount of pseudo-labeling data but gradually increasing the size of labelled set from (\textbf{MCV FLEURS}) to (\textbf{MCV FLEURS Crowd Audiobooks}) subsets.
Relative improvement on MCV  have gradually increased with the size of data from 4\% to 26\% and on FLEURS from 16\% to 36\%. 

These results clearly demonstrate that the ratio of labeled  and unlabeled data has significantly affected the WER improvement, emphasizing its effectiveness in leveraging unlabeled data for low-resource languages. 

\section{Conclusion}

The goals established at the beginning of this project were fully accomplished, significantly advancing the ASR development for underrepresented, low-resource languages such as Armenian. We effectively expanded our datasets and assessed their contributions to the model performance. The key takeaway from our various data extension approaches is that paid crowd-sourcing offers the best balance between cost and quality, outperforming volunteer crowd-sourcing, open-source audiobooks, and unlabeled data. These findings provide actionable insights for optimizing data collection strategies in low-resource settings.

\smallskip

Our top-performing ASR models, trained on these enriched datasets, consistently surpassed existing baselines. Additionally, an ablation study measuring the statistical significance of our findings through POI further confirmed these results. This outstanding performance has prompted us to consider the public release of the models, along with Audiobooks, YouTube source links, and crowd-sourced speech and text data. These data collection approaches have their own effectiveness for enhancing ASR in low-resource languages and costs. These resources, along with the data processing pipelines established in this work, offer a robust foundation for future research.

In conclusion, the data collection presented in this study provide a valuable experience for improving ASR performance in other low-resource languages. By carefully balancing resource investment and output quality, future research can build on our findings to enhance global linguistic inclusivity and accessibility.

\bigskip

\bibliographystyle{IEEEtran}
\bibliography{mybib.bib}

\begin{thebibliography}{10}
\providecommand{\url}[1]{#1}
\csname url@samestyle\endcsname
\providecommand{\newblock}{\relax}
\providecommand{\bibinfo}[2]{#2}
\providecommand{\BIBentrySTDinterwordspacing}{\spaceskip=0pt\relax}
\providecommand{\BIBentryALTinterwordstretchfactor}{4}
\providecommand{\BIBentryALTinterwordspacing}{\spaceskip=\fontdimen2\font plus
\BIBentryALTinterwordstretchfactor\fontdimen3\font minus \fontdimen4\font\relax}
\providecommand{\BIBforeignlanguage}[2]{{%
\expandafter\ifx\csname l@#1\endcsname\relax
\typeout{** WARNING: IEEEtran.bst: No hyphenation pattern has been}%
\typeout{** loaded for the language `#1'. Using the pattern for}%
\typeout{** the default language instead.}%
\else
\language=\csname l@#1\endcsname
\fi
#2}}
\providecommand{\BIBdecl}{\relax}
\BIBdecl

\bibitem{Conneau2022FLEURS:}
A.~Conneau, M.~Ma, S.~Khanuja, Y.~Zhang, V.~Axelrod, S.~Dalmia, J.~Riesa, C.~Rivera, and A.~Bapna, ``Fleurs: Few-shot learning evaluation of universal representations of speech,'' \emph{2022 IEEE Spoken Language Technology Workshop (SLT)}, pp. 798--805, 2022.

\bibitem{Ardila2019CommonVoice}
R.~Ardila, M.~Branson, K.~Davis, M.~Henretty, M.~Kohler, J.~Meyer, R.~Morais, L.~Saunders, F.~M. Tyers, and G.~Weber, ``Common voice: A massively-multilingual speech corpus,'' \emph{arXiv preprint arXiv:1912.06670}, 2019.

\bibitem{bartelds2023making}
M.~Bartelds, N.~San, B.~McDonnell, D.~Jurafsky, and M.~Wieling, ``Making more of little data: Improving low-resource automatic speech recognition using data augmentation,'' \emph{arXiv preprint arXiv:2305.10951}, 2023.

\bibitem{bhogale2024empowering}
K.~S. Bhogale, D.~Mehendale, N.~Parasa, T.~Javed, P.~Kumar, M.~M. Khapra \emph{et~al.}, ``Empowering low-resource language asr via large-scale pseudo labeling,'' \emph{arXiv preprint arXiv:2408.14026}, 2024.

\bibitem{laptev2021dynamic}
A.~Laptev, A.~Andrusenko, I.~Podluzhny, A.~Mitrofanov, I.~Medennikov, and Y.~Matveev, ``Dynamic acoustic unit augmentation with bpe-dropout for low-resource end-to-end speech recognition,'' \emph{Sensors}, vol.~21, no.~9, p. 3063, 2021.

\bibitem{romanenko2022robust}
A.~Romanenko, ``Robust speech recognition for low-resource languages,'' Ph.D. dissertation, Universit{\"a}t Ulm, 2022.

\bibitem{kardava2016georgian}
I.~Kardava, ``Georgian speechrecognizer in famous searching systems and management of software package by voice commands in georgian language,'' in \emph{Conference Proceedings--Third International Conference on Advances in Computing, Electronics and Communication}, vol.~10, 2016, pp. 978--1.

\bibitem{vardanyan2016noise}
A.~Vardanyan, ``Noise-robust speech recognition system for armenian language,'' Ph.D. dissertation, Master’s thesis, American University of Armenia, 2016.

\bibitem{baghdasaryan2022armspeech}
\BIBentryALTinterwordspacing
V.~H. Baghdasaryan, ``Armspeech: Armenian spoken language corpus,'' \emph{International Journal of Scientific Advances}, vol.~3, no.~3, pp. 454--459, 2022. [Online]. Available: \url{https://www.ijscia.com/wp-content/uploads/2022/06/Volume3-Issue3-May-Jun-No.283-454-459.pdf}
\BIBentrySTDinterwordspacing

\bibitem{yeroyan2024enabling}
A.~Yeroyan and N.~Karpov, ``Enabling asr for low-resource languages: A comprehensive dataset creation approach,'' \emph{arXiv preprint arXiv:2406.01446}, 2024.

\bibitem{karpov21_interspeech}
N.~Karpov, A.~Denisenko, and F.~Minkin, ``{Golos: Russian Dataset for Speech Research},'' in \emph{Proc. Interspeech 2021}, 2021, pp. 1419--1423.

\bibitem{stroube2003literary}
B.~Stroube, ``Literary freedom: Project gutenberg,'' \emph{XRDS: Crossroads, The ACM Magazine for Students}, vol.~10, no.~1, pp. 3--3, 2003.

\bibitem{likhomanenko2020slimipl}
T.~Likhomanenko, Q.~Xu, J.~Kahn, G.~Synnaeve, and R.~Collobert, ``slimipl: Language-model-free iterative pseudo-labeling,'' \emph{arXiv preprint arXiv:2010.11524}, 2020.

\bibitem{meister2023librispeech}
A.~Meister, M.~Novikov, N.~Karpov, E.~Bakhturina, V.~Lavrukhin, and B.~Ginsburg, ``Librispeech-pc: Benchmark for evaluation of punctuation and capitalization capabilities of end-to-end asr models,'' in \emph{2023 IEEE Automatic Speech Recognition and Understanding Workshop (ASRU)}.\hskip 1em plus 0.5em minus 0.4em\relax IEEE, 2023, pp. 1--7.

\bibitem{bisani2004bootstrap}
M.~Bisani and H.~Ney, ``Bootstrap estimates for confidence intervals in asr performance evaluation,'' in \emph{2004 IEEE International Conference on Acoustics, Speech, and Signal Processing}, vol.~1.\hskip 1em plus 0.5em minus 0.4em\relax IEEE, 2004, pp. I--409.

\end{thebibliography}

\end{document}